\PassOptionsToPackage{pdfpagelabels=false}{hyperref}
\documentclass[useAMS,usenatbib]{mnras}
\pdfoutput=1 

\usepackage{graphicx}
\usepackage{amsmath,amssymb,amstext}
\usepackage[T1]{fontenc}
\usepackage{ae,aecompl}
\usepackage[utf8]{inputenc}
\usepackage[figure,figure*]{hypcap}
\usepackage[dvipsnames]{xcolor}
\usepackage{bm}
\usepackage{subcaption} 
\usepackage{paralist} 

\usepackage{xparse}

\input{macros.sty}

\usepackage{url}

\PassOptionsToPackage{hyphens}{url}\usepackage{hyperref}  
\hypersetup{
	colorlinks   = true, 
	urlcolor     = blue, 
	linkcolor    = blue, 
	citecolor   = blue 
}

\title[Asymmetric Dark Matter in Stars]{The Effects of Asymmetric Dark Matter on Stellar Evolution I: Spin-Dependent Scattering}

\author[T.J. Raen et al.]{%
Troy J. Raen,$^{1}$\thanks{E-mail: troy.raen@pitt.edu},
Héctor Martínez-Rodríguez$^{1}$, 
Travis J. Hurst$^{2}$,
Andrew R. Zentner$^{1}$,\newauthor
and Carles Badenes$^{1}$,
and Rachel Tao$^{3}$
\vspace*{12pt}
\\
$^{1}$Department of Physics and Astronomy \& Pittsburgh Particle Physics, Astrophysics, and Cosmology Center (Pitt PACC),\\ University of Pittsburgh, Pittsburgh, PA 15260, USA\\
$^{2}$Department of Mathematics and Physics, Colorado State University - Pueblo, Pueblo, CO, 81001 \\
$^{3}$Department of Physics, Emory University, Atlanta, GA 30322
}
\date{\today}

\pubyear{2020}

\begin{document}
\label{firstpage}
\pagerange{\pageref{firstpage}--\pageref{lastpage}}
\maketitle

\begin{abstract}
Most of the dark matter (DM) search over the last few decades has focused on WIMPs, but the viable parameter space is quickly shrinking. Asymmetric Dark Matter (ADM) is a WIMP-like DM candidate with slightly smaller masses and no present day annihilation, meaning that stars can capture and build up large quantities. The captured ADM can transport energy through a significant volume of the star. We investigate the effects of spin-dependent ADM energy transport on stellar structure and evolution in stars with \mrange in varying DM environments. We wrote a MESA module\footnotemark\xspace that calculates the capture of DM and the subsequent energy transport within the star. We fix the DM mass to 5 GeV and the cross section to $10^{-37}\ \mathrm{cm{^2}}$, and study varying environments by scaling the DM capture rate. For stars with radiative cores (\mrangelow), the presence of ADM flattens the temperature and burning profiles in the core and increases MS ($X_c > 10^{-3}$) lifetimes by up to $\sim 20\%$. 
We find that strict requirements on energy conservation are crucial to the simulation of ADM's effects on these stars. 
In higher-mass stars, 
ADM energy transport shuts off core convection, limiting available fuel and shortening MS lifetimes by up to $\sim 40\%$. This may translate to changes in the luminosity and effective temperature of the MS turnoff in population isochrones.
The tip of the red giant branch may occur at lower luminosities. 
The effects are largest in DM environments with high densities and/or low velocity dispersions, making dwarf and early forming galaxies most likely to display the effects.
\end{abstract}

\footnotetext{\url{https://github.com/troyraen/DM-in-Stars}}




\section{Introduction}
\label{sec:intro}

  A preponderance of the evidence suggests that approximately $84\%$ of the matter budget of the 
  universe consists of a form of non-baryonic dark matter that has yet to be identified 
  \citep[e.g.,][]{jungman_etal96,Bertone+05,CosmicVisions17,Profumo+19}. 
  In the standard picture of cosmological structure formation, 
  galaxies form within the potential wells of large, 
  nearly virialized halos of dark matter \citep{white_rees78,blumenthal_etal84}. 
  If the dark matter interacts with standard model particles, 
  it can be captured by stars moving through dark matter halos 
  \citep{press_spergel85,krauss_etal85,gaisser_etal86,griest_seckel87}. 
  Once captured, continued scattering within the stellar interior contributes 
  to energy transport within the star, potentially altering its evolution \citep{Spergel1985EffectInterior,Taoso+10,Zentner2011AsymmetricDwarfs,Iocco+12,Lopes_Silk12,Casanellas_Lopes13,Casanellas+15,vincent_etal15,Murase_Shoemaker16,Lopes_silk19,Vincent2020}. 
  The significance of this energy transport depends on the following 
  properties (in addition to the properties of the star): 
  (1) the DM mass, $\mx$; 
  (2) the DM-nucleon scattering cross section, $\sigxn$; 
  and (3) the total number of DM particles captured by a star, $\Nx$, 
  which itself depends on $\mx$ and $\sigxn$ as well as the local DM environment from 
which the particles are captured (see \S~\ref{sec:props}). 
We study the effects of energy transport by asymmetric dark matter 
(ADM, see below)
in stars of mass \mrange living within a variety of dark matter 
environments using the publicly-available code 
Modules for Experiments in Stellar Astrophysics 
\citep[\mesa,][]{Paxton2011ModulesMESA, Paxton2013, Paxton2015, Paxton2018, Paxton2019}.

  Evidence supporting the claim that $\sim$84\% of the matter in the universe is in 
  an unknown form of dark matter is abundant and varied, ranging from the 
  anisotropy of the microwave background radiation to formation and structures of galaxies 
  \citep[e.g.,][]{jungman_etal96,Bertone+05,PLanck18}. 
  For several decades, the leading candidate has been the so called 
  Weakly-interacting massive particle (WIMP). 
  The classic WIMP is a heavy ($\mx \sim 10^2-10^3 \gev$) 
  thermal relic whose contemporary abundance is set 
  by its annihilation rate in the early universe 
  \cite[e.g.,][]{kolb_turner90}. 
  Therefore, WIMPs are thought to have a fairly well established ``standard'' annihilation 
  cross section \citep[e.g.,][]{steigman_etal12}, which is comparable to typical weak-scale 
  cross sections, $\langle \sigma v \rangle \sim 10^{-26} \, \mathrm{cm}^3/\mathrm{s}$. 
  This annihilation of WIMPs, which is so critical to guaranteeing that the 
  correct abundance of dark matter in the contemporary Universe, in turn, 
  limits the number of particles that can accumulate within a star. 
  The rate of capture of new dark matter particles 
  comes to equilibrium with dark matter particle annihilation in the 
  stellar interior \citep{krauss_etal85}. 
  Despite numerous ongoing terrestrial direct detection experiments 
  \citep[see][for a review]{Schumann19} 
  and efforts to detect dark matter indirectly through 
  its annihilation products \citep[reviewed in][]{Slatyer17}, 
  dark matter has not been observed non-gravitationally. The 
  available parameter space for relatively light ($\mx \lesssim 10^2 \mathrm{GeV}$) DM
  is rapidly shrinking, which has triggered a surge in research into 
  alternatives to the long-favored WIMP.

  Asymmetric dark matter (ADM) is an alternative to the classic WIMP in which 
  the relic abundance of the dark matter particle is set by a primordial asymmetry 
  rather than via annihilation \citep[for a review, see][and references therein]{adm_review,Petraki2013wwa}. 
  If the baryon and dark matter asymmetries are 
  related, then such models have the appealing property that they explain 
  the fact that the contemporary dark matter and baryon abundances are 
  of the same order of magnitude, which is otherwise surprising because 
  these relic abundances are determined by unrelated physics in the WIMP 
  scenario. Indeed, this was one of the early motivations for ADM-like 
  models \citep[e.g.,][]{Nussinov85,barr_etal90,chivukula_walker90,kaplan92}. 
  The variety of specific incarnations of ADM is broad, 
  but ADM models often predict particle masses smaller than 
  the classic WIMP ($\mx \sim 1-10 \gev$) and little or no 
  contemporary dark matter annihilation for lack of relic 
  dark matter anti-particles.

  These predictions motivate studies to 
  constrain ADM indirectly through stellar astrophysics. The lack of 
  annihilation means that ADM may build up to very large
  quantities within stars because the capture of ADM is never countered 
  by annihilation. Meanwhile, the relatively low masses 
  compared to the classic WIMP mean captured ADM particles orbit within 
  a significant volume of the star, out to $\rx \sim 0.1 R_{\odot}$
  for a Sun-like star, which means that they experience large differences 
  in ambient temperature 
  throughout their orbits and can thus transport energy outward from the 
  stellar core extremely efficiently \citep{Spergel1985EffectInterior}. 
  These features of ADM have already motivated research into the possibility that 
  ADM may alter stellar evolution 
  \citep[e.g.][]{Taoso+10,Zentner2011AsymmetricDwarfs,Iocco+12,Lopes_Silk12,Casanellas_Lopes13,Casanellas+15,vincent_etal15,Murase_Shoemaker16,Lopes_silk19,Lopes_2019,Vincent2020}. Our results are generally in agreement with previous works, insofar as they can be compared considering variations in the chosen parameters of each study which can include ADM properties (e.g., mass and cross section), halo environments (e.g., ADM density and velocity dispersion in the stellar neighborhood), and stellar mass.
  In this paper we undertake a study of the properties and evolution of 
  stellar populations within halos of ADM. We fix the ADM mass and cross-section and study the effects of varying halo environments across a wider range of parameter space than has been done previously. (See \S~\ref{sec:props} for a discussion of ADM properties and environments.)
  The effects of stellar cooling are particularly large in environments in which the ambient dark matter density is high and velocity dispersion is low, such that the capture of dark matter is extremely efficient. Thus, these effects will be largest in dwarf satellite galaxies and high-redshift galaxies.   
  In this first paper on the topic, 
  we further narrow our study to spin-dependent ADM-nucleus scattering. 
  Spin-independent ADM-nucleus scattering leads to behaviors that are 
  qualitatively distinct from spin-dependent scattering; therefore, we will 
  present results for the former case in a forthcoming manuscript. 
  
  In the following section, we summarize the dependence of the capture rate of dark matter 
  within stars on both dark matter and stellar properties. In \S~\ref{sec:methods}, we describe
  our simulations of stellar evolution including cooling due to ADM. We present our results 
  in \S~\ref{sec:results}. We discuss our results and draw conclusions in \S~\ref{sec:discus}.

\section{Dark Matter Properties and Capture in Stars}
\label{sec:props}

  Probing the parameter space of ADM with simulations 
  of stellar evolution is computationally expensive. 
  Consequently, we show results for an illustrative set of 
  ADM parameters that we initially chose in order to: 
  (1) make the effects of ADM on stellar evolution significant; 
  (2) narrowly-evade the evaporation threshold; and 
  (3) remain consistent with contemporary constraints on 
  dark matter properties (but see the discussion in the next 
  paragraph). We choose $\mx = 5 \gev$ and a 
  spin-dependent dark matter--nucleon scattering cross section of 
  $\sigxp = 10^{-37} \cmsq$. Hereafter, we will discuss ADM-proton 
  scattering since protons are the only nuclei in MS stars with both 
  a significant abundance and a net spin.
  With these parameters, ADM evaporation is negligible. 
  We find that the largest evaporation mass in our models is 
  $\simeq 4.6 \gev$ at a solar mass of $\simeq 1.4 \Msun$, 
  consistent with the literature on this topic (see, for example, 
  the classic papers by \citealt{Gould87} and \citealt{1990ApJ...356..302G} 
  and the recent update by \citealt{2017JCAP...10..037B}). 
  We note that evaporation masses are slightly higher for 
  spin-independent scattering, because the helium nucleus is 
  more closely matched to the mass of the dark matter particle, 
  so this must be considered in any extension of this work 
  to spin-independent interactions. 
  We assume that ADM self-interactions 
  are negligible throughout; however, 
  it is likely that self-interactions, if they existed, 
  would lead to enhanced cooling 
  \citep[e.g.,][]{Zentner2009High-energySun}. 
  Exploring such models 
  would constitute a potentially 
  interesting follow-up to this work.

  During the course of this 
  work, the PICO collaboration was able to reduce its thresholds, 
  unambiguously excluding dark matter with the specific parameter 
  values listed above \citep{PICO}. While the simulations 
  that we have carried out are computationally intensive and it is 
  impractical to repeat each of the $\sim 600$ stellar evolution 
  simulations, we believe our results are a meaningful exploration of the deviations in stellar evolution due to ADM and our energy transport module provides a basis for further exploration in future work. Consequently, we choose to present these results 
  as a qualitative indication of the effects that ADM can have 
  upon stellar evolution. As we will see below, there is 
  significant uncertainty involved in associating a particular 
  stellar effect with a particular dark matter-nucleon scattering 
  cross section due to a variety of model uncertainties. 
  Consequently, while the effects of dark matter within stars 
  will likely be milder than those that we describe here, it is 
  possible that the same, qualitative effects could be realized 
  in nature.

The amount of energy transported by dark matter 
(see \S~\ref{sub:energytransport}) is proportional 
to the amount of ADM within the star. 
In ADM models, in which annihilation of dark matter is negligible, 
the number of dark matter particles within the star at any given time, 
$t$, is determined by ${\rm d}\Nx/{\rm d}t = \Cx$, 
where $\Cx$ is the instantaneous ADM capture rate. 
We use the capture rate from \citet{Zentner2011AsymmetricDwarfs}, 
which is a simplified form valid for dark matter particle masses 
$\mx \lesssim 20 \gev$ \citep[see][for more complete capture rates]{Gould1992CosmologicalAnnihilations,Zentner2009High-energySun}:
  \begin{align}
  \begin{split}
    \label{eq:capturerate}
    \Cx =\ & \Csun
    \Big(\frac{\rhox}{0.4 \gev \cmcinv}\Big)
    \Big(\frac{270 \kms}{\bar{v}}\Big) \\
    & \times \Big(\frac{\sigxp}{10^{-43} \cmsq}\Big) \Big(\frac{5 \gev}{\mx}\Big) \\
    & \times \Big(\frac{v_{\mathrm{esc}}}{618 \kms}\Big)^2
    \Big(\frac{\Mstar}{\Msun}\Big)
  \end{split}
  \end{align}
  where $\Csun = 5 \times 10^{21} \sinv$, 
  $\rhox$ is the DM density in the stellar environment,
  $\bar{v}$ is the velocity dispersion of dark matter particles 
  in the stellar neighborhood, and $v_{\mathrm{esc}}$ is the escape speed from the 
  surface of the star.

  The first line of Eq.~(\ref{eq:capturerate}) gives 
  the dependence of the capture rate on the stellar environment. 
  Both $\rhox$ and $\bar{v}$ are properties of the local stellar environment 
  and are degenerate with one another in Eq.~(\ref{eq:capturerate}). 
  A higher ambient density of dark matter leads to a higher rate of dark matter capture, 
  while a lower relative velocity between the star and the infalling dark matter leads 
  to a higher probability for capture. 
  Because of this degeneracy, coupled with the fact that these parameters carry 
  considerable uncertainty themselves, 
  it is convenient to parameterize a star's 
  local dark matter environment by an overall 
  factor \citep{Zentner2011AsymmetricDwarfs,Hurst2015},
  \begin{equation}
  \gb = \bigg(\frac{\rhox}{0.4 \gev \cmcinv}\bigg) \Bigg(\frac{270 \kms}{\bar{v}}\Bigg).
  \label{eq:gammab}
  \end{equation}
%
Normalized in this way, $\gb$ specifies the capture rate, 
$\Cx$, relative to the rate that would be realized in the 
solar neighborhood for the same star. 
From this point on, 
we will characterize a star's dark matter environment using 
$\gb$. In general, we will be most interested in values of $\gb > 1$. 
A value of $\gbzero$ describes a stellar environment with no dark matter 
(hereafter referred to as `standard models' and labeled `\nodm'), 
and $\gbone$ describes the solar neighborhood. 
A value of $\gbpow{2}$ may specify an environment in 
which the dark matter density is 100 times that in the 
solar neighborhood at the same velocity dispersion, 
an environment in which the velocity dispersion is 1/100 that 
of the solar neighborhood at the same density, 
or any of an infinite number of other possible combinations.

It is interesting to consider the range of $\gb$ values 
that would be considered reasonable. If the distribution of 
dark matter within galaxies, such as the Milky Way, follows a 
profile that diverges as the 
\citet[][NFW]{nfwprofile} density profile, then one might expect 
to find dark matter environments near the centers of galaxies with
densities significantly higher than the local value and velocity dispersions significantly
lower than the local value, giving $\gb \gg 1$. Such scenarios 
were explored in \citet{Bertone_Fairbairn08}, \citet{Fairbairn+08}, 
and \citet{Scott+09}. While such large values of $\gb$ may well 
lead to large effects on stellar structure, 
stellar populations near the Galactic Center
are difficult to observe and any assumption 
about the dark matter density profile in the 
inner regions of any galaxy must be considered speculative. 
Interestingly, Local Group dwarf galaxies are extremely dark matter-dominated 
and have well-constrained dark matter profiles and velocity dispersions. 
In some cases, the Local Group dwarfs have densities $\sim 3$ orders 
of magnitude higher than the dark matter density in the Solar neighborhood 
and velocity dispersions that are at least $\sim 2$ orders of magnitude 
smaller than the local value. This suggests that values of $\gb \sim 10^{5-6}$ 
could be realized within Local Group dwarf galaxies and has the further merit that 
$\gb$ within Local Group dwarfs can be measured more precisely in the future. 
A third possibility for large values of the environmental factor 
are early-forming, very high-redshift galaxies. These galaxies begin 
forming in small, dense halos where the environmental boost factor 
can reach $\gb \gtrsim 10^6$ at redshifts $z \gtrsim 10$ \citep{KBD}. 
Of course, these stars will not be directly observable, but it 
is interesting to speculate that such stars could be detected as 
remnants of early mergers with the proto-Milky Way and/or that 
changes to the structure and evolution of these stars could be detected 
indirectly in the chemical evolution of the larger, lower-redshift 
galaxies in which they will be found today.

Finally, while we have focused on the environmental parameter, $\gb$, 
as a proxy for the dark matter environment in which a star is embedded, 
we note that values of $\gb \ne 1$ can also be mimicked through dark matter 
physics. In particular, dark matter self-interactions can greatly 
enhance the capture rates of dark matter within 
stars \citep{Zentner2009High-energySun}. This effect of dark matter 
self-capture itself grows with increasing ambient density and 
decreasing ambient velocity dispersion, so the two effects 
reinforce one another. For example, a value of $\gb \sim 10^4$ may 
be realized by increasing the ambient dark matter density by a 
factor of $\sim 10^3$, while simultaneously introducing a 
dark matter self-interaction that boosts the number of 
captured dark matter particles by a factor of $\sim 10$. 
We relegate the separation of these effects to future work 
and encapsulate all of this uncertainty into the single 
parameter $\gb$.

\section{Methods}
\label{sec:methods}

We study the impact of dark matter on the evolution of \mrange stars (with a mass step of $0.05 \Msun$) through core helium depletion ($Y_c = 10^{-3}$) or a maximum age of 10 Gyr, whichever comes first, using the publicly-available code 
Modules for Experiments in Stellar Astrophysics \citep[\mesa,][]{Paxton2011ModulesMESA, Paxton2013, Paxton2015, Paxton2018, Paxton2019}, 
release $12115$. 
We used the \texttt{MESA SDK} version $20190830$\footnote{\url{https://zenodo.org/record/3560834}} to compile \mesa.
We base our stellar parameter inlist on the \mesa test suite \texttt{1M\_pre\_ms\_to\_wd} inlists and use a metallicity of $Z = 0.0142$. Twelve of the models we ran did not complete (e.g., due to requiring unreasonably small timesteps), and we have excluded them from our final data set. Of these, 2 were $\gbpow{4}$ models, and none of them were either \nodm or $\gbpow{6}$ models (these are the three $\gb$ values we highlight below). We wrote a module that calculates dark matter capture and energy transport (see \S~\ref{sub:energytransport}) and connects to \mesa simulations via the provided \texttt{extra\_energy\_implicit} hook. For the reader interested in either examining the underlying data or reproducing our results, we make the following \mesa input/output files available on 
Zenodo\footnote{\url{https://zenodo.org/record/4064115}}
and through the 
\mesa Marketplace\footnote{\url{http://cococubed.asu.edu/mesa_market/add-ons.html}}:
\begin{inparaenum}[1)]
\item inlist templates and \texttt{src} files;
\item $1.0 \Msun$ and $3.5 \Msun$ data for \nodm, $\gbpow{4}$, and $\gbpow{6}$ models.
\end{inparaenum}
Additional data will be shared on reasonable request to the corresponding author.
For the reader interested in utilizing our module to explore the effects of varying parameters beyond the scope of this paper, our code is available on GitHub\footnote{\url{https://github.com/troyraen/DM-in-Stars}}.
 
We take advantage of the significantly improved numerical energy conservation capabilities in recent \mesa versions \citep[introduced in][]{Paxton2019} as we find this to be crucial to a proper accounting of the effects of ADM. Energy transport by significant amounts of AMD alters the core structure of a star such that small changes in the temperature profile due to poor energy conservation can lead to a reversal in the direction of ADM energy transport which destabilizes the star. See \S~\ref{sub:lowmass} for further discussion.



To generate isochrones (see \S~\ref{sub:isochrones}) from \mesa's stellar models we perform a linear interpolation of our stellar tracks to a uniform set of ages and choose isochrones that are as well-sampled as possible in regions of interest. \mesa's adaptive time steps resolve dynamic phases of evolution quite well, and this interpolation is not problematic. 
We also show one isochrone generated from the \mist code \citep[\mesa Isochrones and Stellar Tracks,][]{Dotter2016MesaIsochrones, Choi2016MESAModels}, which uses a multi-step process to interpolate both the stellar tracks and the mass grid and therefore resolves otherwise sparsely populated regions of the isochrones. We are unable to show more \mist isochrones because the interpolation failed in most of our parameter space; most notably, it did not produce any isochrones older than 1 Gyr. This is likely due to non-monotonicity in the mass-age relation at fixed evolutionary phase (see \S~\ref{sub:mstau}) which violates assumptions of the code.

\subsection{Energy Transport by Dark Matter}
\label{sub:energytransport}

  The energy transported by captured ADM can, in principle, be computed by solving the Boltzmann equation; however, this strategy is too computationally intensive to combine with a full-scale simulation of the evolution of stellar structure. To reduce the computational costs of our simulations, we estimate ADM energy transport using the approximations of \citet{Spergel1985EffectInterior}. In particular, we assume a Maxwellian phase-space distribution for the ADM and calculate an orbit-averaged temperature, $\Tx$, by requiring that the distribution satisfy the first moment of the Boltzmann equation. This amounts to a requirement on energy conservation: ADM should neither inject nor remove a net energy from the star. The rate of energy transfer (per unit mass) from dark matter to protons is then
  %
  \begin{align}
  \begin{split}
  \epsx(r) =\ & 8\ \sqrt[]{\frac{2}{\pi}} \frac{\nx(r) \nprot(r)}{\rho(r)} \frac{\mx \mprot}{(\mx+\mprot)^2} \sigxp\\
  & \times \left(\frac{\mprot k \Tx + \mx k T(r)}{\mx \mprot}\right)^{1/2} k [\Tx - T(r)],
  \label{eq:xheat}
  \end{split}
  \end{align}
  where $n(r)$ is a number density, $\rho(r)$ is the mass density, $k$ is Boltzmann's constant, and the subscript p refers to protons. \citep[See][for a detailed derivation]{Spergel1985EffectInterior}.

Generally, $\nprot$, $\nx$, and $T$ all peak at the center, so the energy transport is most efficient here. The number density 
of dark matter particles, $\nx$ increases in proportion with $\Nx$, so we can expect the effects to increase with both $\gb$ and stellar age through the MS, while hydrogen is abundant. As a star leaves the MS, $\nprot$ drops in the core and spin-dependent ADM energy transport is greatly diminished because there are relatively few protons left with which dark matter may scatter\footnote{This is one of the primary reasons that spin-dependent and spin-independent scattering give qualitatively distinct results. As the star burns H on the MS, the number of protons is reduced, reducing the importance of spin-dependent scattering processes. In the case of spin-independent scattering, the effect gets more important as helium 
is produced from H burning during the MS.}. 

  The sign of $\epsx(r)$ is given by the final term in (\ref{eq:xheat}), $\Tx - T(r)$, which is used to define an ADM characteristic radius, $\rx$, implicitly as
  \begin{align}
    T(r=\rx) = \Tx.
  \end{align}
  Then dark matter takes energy from $r < \rx$ and deposits it at $r > \rx$ for a standard MS temperature profile (monotonically decreasing from the center outward). With our chosen ADM parameters we see typical values:
  \begin{align}
    \rx & \sim \mathcal{O}(0.1 \Rstar) \\
    \lx = (\sigxp \nprot)^{-1} & \sim\mathcal{O}(1 \Rstar)
  \end{align}
  where $\lx$ is the ADM mean free path (implying that it completes several orbits between scattering events). These values allow dark matter to travel much larger distances than photons or ions within the star (which have $l \lesssim 10^{-10} \Rstar$) and to traverse qualitatively distinct 
  regions of the star. This large mean free path is what enables dark matter 
  to serve as such an effective coolant despite being far less numerous than either photons or ions \citep{Spergel1985EffectInterior}.


\section{Results}
\label{sec:results}

In standard stellar evolution, with no influence from dark matter, stars with \mrange naturally split into two groups with qualitatively different structures, based on the dominant channel through which they burn hydrogen. Spin-dependent asymmetric dark mater (ADM) affects core hydrogen burning, mainly by flattening the temperature gradient. In this section, we will review standard stellar astrophysics \citep{Kippenhahn2012} and then describe the effects of asymmetric dark matter seen in our MESA simulations.

The dominant burning channel is determined by the core temperature, 
with the transition happening at $\Tc \sim 2 \times 10^7 \K$, 
which corresponds to a stellar mass of $\Mstar \sim1.3 \Msun$.
Stars with $\Mstar \lesssim 1.3 \, \Msun$, which we call low-mass stars, 
burn hydrogen primarily through the proton-proton (pp) chain 
for which the burning rate scales with temperature very roughly as $\epspp \propto T^4$. 
For stars in the mass range $0.4 \lesssim \Mstar/\Msun \lesssim 1.3$, 
the transport of energy away from the core burning region is dominated 
by photon diffusion. Energy transport in the cores of such stars is said 
to be radiative.

Stars with $\Mstar \gtrsim 1.3 \Msun$, 
which we call high-mass stars, 
are dominated by the carbon-nitrogen-oxygen (CNO) cycle, 
for which the burning rate scales much more strongly with temperature, 
$\epsCNO \propto T^{16-20}$. In CNO-dominated stars, 
radiative energy transport is insufficient 
to carry away the energy produced by hydrogen burning; consequently, they have convective cores.

In \S~\ref{sub:lowmass} and \S~\ref{sub:highmass}, we will consider results for low-mass stars and high-mass stars separately and we will demonstrate that ADM has distinct effects on the evolution of the two groups. \S~\ref{sub:mstau} details changes in MS lifetimes due to ADM energy transport. We discuss the effects on surface properties of individual stars in \S~\ref{sub:tracks}, and on the isochrones of stellar populations in \S~\ref{sub:isochrones}.
Note that all logarithms in this paper are base 10.

\subsection{Low-Mass Stars}
\label{sub:lowmass}

  \begin{figure}
    \centering
    \includegraphics[width=0.47\textwidth]{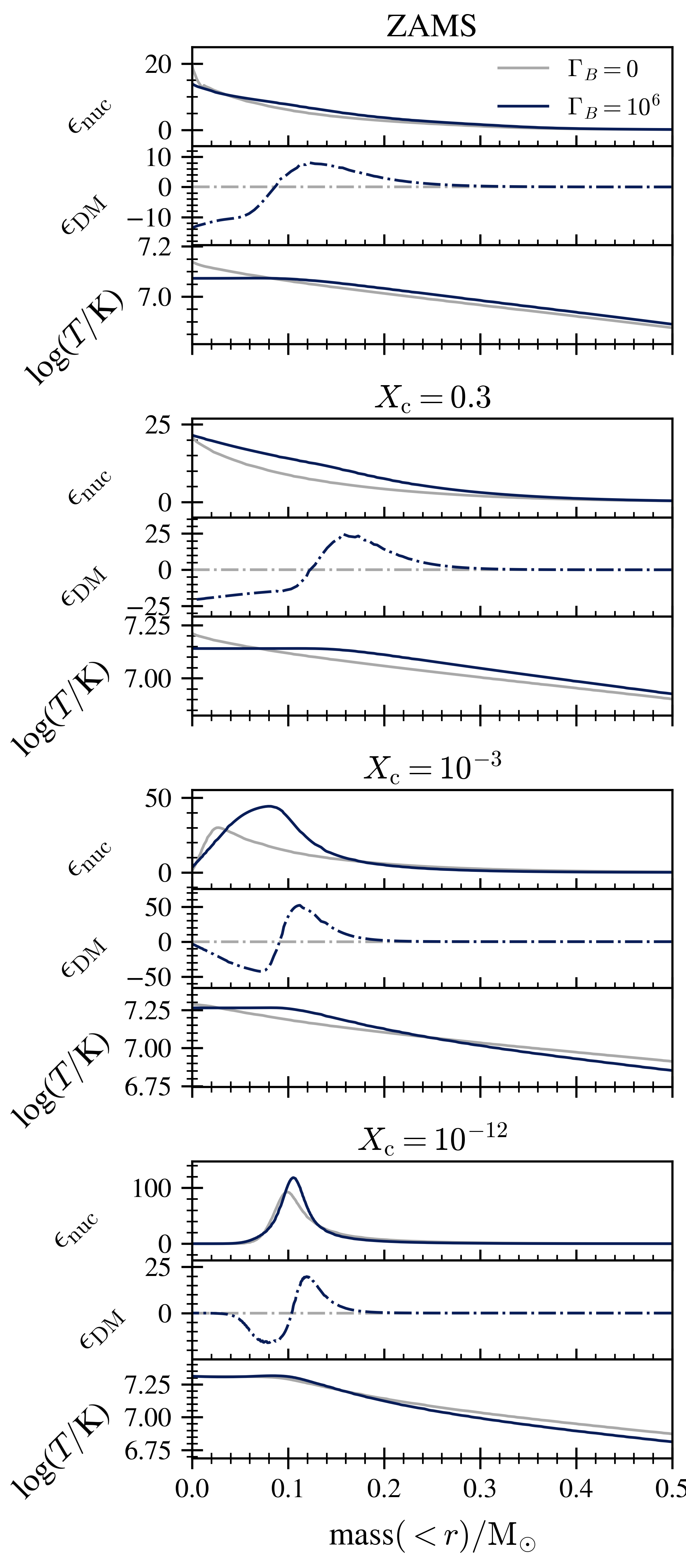}
    \caption{$1.0 \Msun$ profiles for \nodm (grey) and $\gbpow{6}$ (dark blue) models. Each set of 3 panels shows stellar profiles of the stars at different evolutionary phases indicated by the fraction of hydrogen in the center, $X_c$, which decreases as the star evolves (ZAMS is "zero-age main sequence"). The profiles in each panel are: 1) $\epsnuc$, the nuclear burning rate in [erg/g/s]; 2) $\epsx$, the rate at which DM transports energy (negative values indicate that energy is being removed), also in [erg/g/s]; 3) log($T$/K), log$_{10}$ of the temperature in [K]. ADM energy transport decreases the temperature and burning rate in the center and increases them in a shell at $m(<r) \sim 0.1 \Msun$. 
    }
    \label{fig:m1p0c6}
  \end{figure}

Standard model stars in the mass range \mrangelow have relatively 
low central temperatures and so are powered primarily by the pp chain, 
which is much less sensitive to the temperature than burning via the CNO cycle. 
This means the burning does not peak as strongly at the center and 
radiative transport is sufficient to carry the energy flux, so the core is radiative. 
Without the mixing provided by convection, 
hydrogen depletes first at the very center 
and the burning shifts gradually outward into a shell.

As seen in Figure~\ref{fig:m1p0c6}, 
energy transport by large amounts of ADM causes flatter temperature gradients in the center 
than those seen in the \nodm model. This reduces the burning rate 
in the center (as long as the local density does not get too high), where ADM is removing energy, and increases 
it in a shell, where ADM deposits energy.
Note that in Figure~\ref{fig:m1p0c6} the burning rate is not significantly reduced at the center; this is due to a significant increase in density (with which the burning rate per gram scales linearly) as the star's structure adjusts to cooling in the core. 
These results are generally in agreement with the results of previous papers studying ADM with similar properties.
\cite{Taoso+10} found decreased core temperatures in the models of the sun affected by spin dependent ADM with $\sigxp = [1, 2, 3] \times 10^{-36} \cmsq$ and $\mx = 7 \gev$. 
\cite{Iocco+12} found similarly altered temperature and burning profiles in solar mass models affected by ADM with $\sigxp = 10^{-37} \cmsq$ and $\mx = 10 \gev$ in $\sim \gbpow{3}$ environments. Our work uses similar or slightly lower values and considers the full range of environments expected to exist in nature.

The increased burning at 
larger radii ($m(<r) \sim 0.1 \, \Msun$) 
causes a small net increase in the total 
luminosity of the star. Because ADM probes temperature differences 
over large portions of the star, 
energy transport by ADM remains efficient despite the shallower 
temperature gradient. The increased temperature at $m(<r) \gtrsim 0.1 \Msun$ 
means that more hydrogen burns during the MS in low-mass ADM models 
than in their \nodm counterparts. These two competing effects 
dictate the gross evolution of the star. For low values of 
$\gb$ (weaker ADM influence), the increase in the total 
amount of fuel wins and the stellar MS lifetime is an  
increasing function of $\gb$. At values of $\gb \gtrsim 10^3$, 
the burning rate continues to increase, but little additional 
fuel is burned because of the precipitous drop in temperature 
at larger radii, which cannot be overcome by the energy transport. 
The result is that the increase in MS lifetimes peaks at $\gb \sim 10^3$ after which MS lifetimes decrease with increasing environmental factor, eventually declining below the lifetimes of \nodm stars at a point which depends on both the stellar mass and $\gb$. 
(We will discuss stellar lifetimes further in \S~\ref{sub:mstau}).
Stars in this regime have a higher surface luminosity than their \nodm counterparts, at fixed central hydrogen fraction. However, at fixed luminosity we should expect their effective temperatures to be roughly the same, because these are equilibrium points dictated by the equations of stellar structure. Indeed, this is what we see in our models. The result is that stars of a given mass but different ADM content move along roughly the same tracks in an HR diagram, but they do so at different rates. (We will discuss stellar tracks in \S~\ref{sub:tracks} and population isochrones in \S~\ref{sub:isochrones}.)

In our initial simulations we used an earlier version of \mesa (release $10398$) which included less rigorous energy conservation requirements and produced very different results in this low-mass regime. In those models, energy transport by large amounts of ADM caused the temperature profile 
to become inverted, with the central temperature falling below the ADM temperature. Once this occurred, ADM energy transport reversed direction and began moving energy {\em toward the center} of the star. This raised the central temperature until it surpassed the ADM temperature again, 
causing ADM energy transport to again reverse direction and move energy {\em away} from the center. 
This cycle was self-reinforcing and resulted in large oscillations in the core temperature, density, etc., which propagated outward and resulted in large 
oscillations in surface properties as well.
Previous work by \cite{Iocco+12} found similar dramatic oscillations in solar mass stars 
in dense ADM environments, 
and noted that they were unable to determine whether it was 
a physical effect or a numerical artifact. 
Upon further investigation of our initial models, 
we found that they had poor energy conservation. 
A new \mesa version had been released since we had begun 
this work that included improved energy conservation schemes. 
When we updated to \mesa release $12115$ and ran the models again, 
the energy conservation was much improved, 
and the central temperature was reduced such that 
it was very close to the ADM temperature but never 
dropped below it. Since $\Tc > \Tx$ throughout the star's lifetime, 
ADM energy transport never reverses direction and the oscillations 
seen previously are now absent. We conclude that the oscillations in our initial simulations
were a numerical artifact, and that strict energy conservation requirements are necessary for a proper accounting of ADM effects.

\subsection{High-Mass Stars}
\label{sub:highmass}

  \begin{figure}
    \centering
    \includegraphics[width=0.47\textwidth]{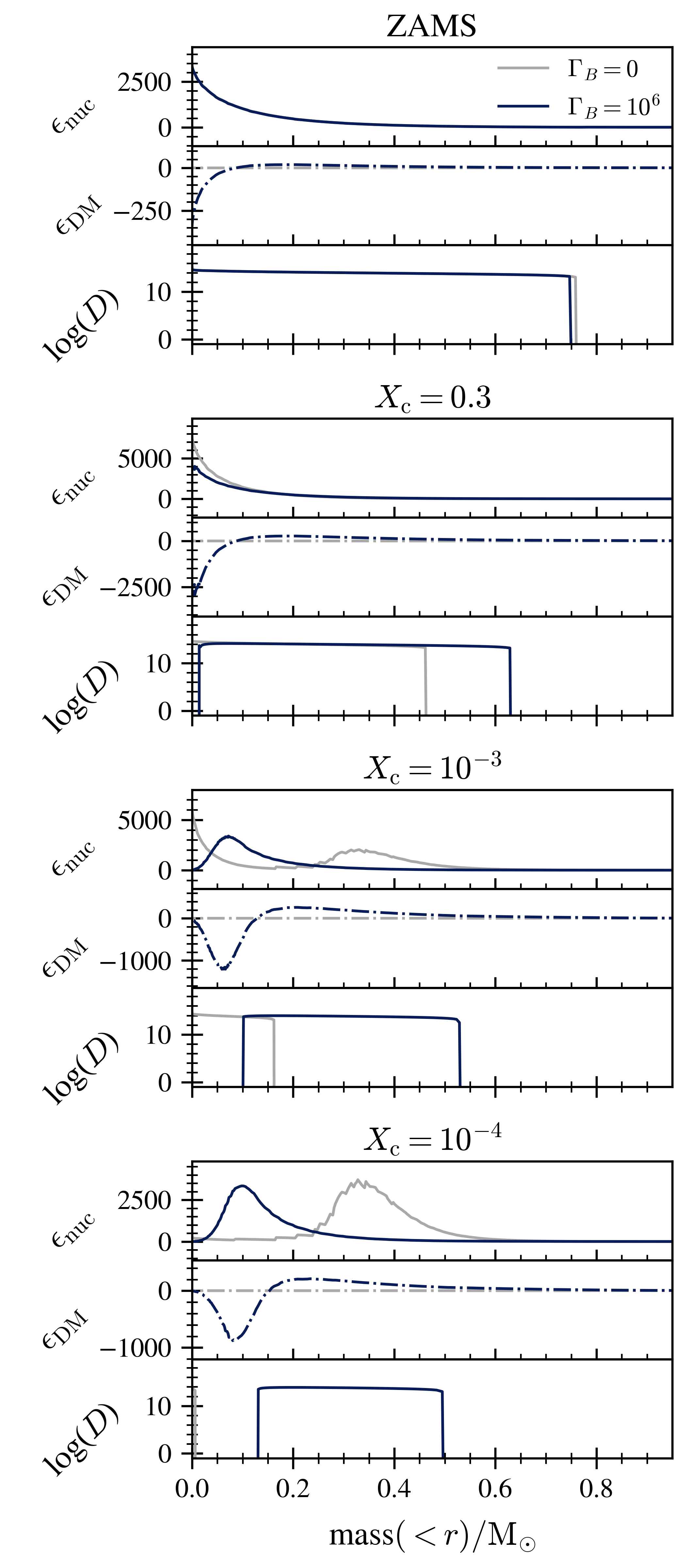}
    \caption{Same as Figure~\ref{fig:m1p0c6} for $3.5 \Msun$ models, except that the 3rd panel in each set shows log($D$), where $D$ is the diffusion coefficient for convective mixing in [cm$^2$/s]. In the \nodm model the convective core retreats \textit{toward} the center over time, and the burning rate peaks at the center until the end of the main sequence when the burning rate drops dramatically and a shell of strong burning appears suddenly. In the $\gbpow{6}$ model, convection at the very center shuts off relatively early in the MS and a convective shell retreats \textit{away} from the center over time. The peak burning rate shifts gradually outward, following the inner edge of the convective shell. The $\gbpow{6}$ model reaches the $X_c$ evolutionary markers at younger ages, relative to the \nodm model, since convection cannot replenish the fuel at the center.
    }
    \label{fig:m3p5}
  \end{figure}

In standard models, MS stars with $\Mstar \gtrsim 1.3 \Msun$ 
are powered primarily by the CNO cycle.
This has several important consequences:
(1) the burning rate is much higher than in pp-dominated stars;
(2) the burning rate is extremely sensitive to core temperature; 
(3) temperature gradients in the stellar core are relatively steep; 
and (4) stellar cores must be convective in order 
to carry away the energy produced by core hydrogen burning. 
Convective energy transport in the star also replenishes the 
core with unburnt hydrogen as the star evolves.
Once hydrogen throughout the convective zone is depleted, 
the burning rate rapidly decreases and 
the star loses more energy at its surface than is 
being generated by burning. Gravity temporarily overcomes pressure support and 
the star contracts until the internal temperature increase is sufficient 
to ignite hydrogen in a shell outside the depleted core. This restructuring 
produces the so-called "convective hook" in an HR diagram 
as the star leaves the main sequence \citep{Kippenhahn2012}.

If a star captures enough ADM, the combination of dark matter + radiative energy 
transport becomes sufficient to carry the flux from nuclear burning at a 
temperature gradient that is insufficient to support convection. 
In other words, the additional energy transport 
by ADM can turn off convection within the stellar core. 
This can be seen in 
Figure~\ref{fig:m3p5} for a $3.5 \Msun$, $\gbpow{6}$ star. 
Convection disappears from the center first 
(where ADM energy transport is most efficient) 
and retreats away from the core, into a narrowing shell. 
Without convective mixing, the hydrogen fuel supply depletes first 
at the very center (instead of simultaneously throughout the core) and 
the burning concomitantly shifts gradually into a shell, 
following the lower boundary of the convective zone. 
This can be seen in the time progression (down the page) 
of the $\gbpow{6}$ (dark blue) model in Figure~\ref{fig:m3p5}. 
The shift to shell burning is similar to the behavior of standard {\em low mass} 
stars. 

The suppression of convection in the cores of stars affected by ADM has been noted in previous work. 
It was first predicted to occur in horizontal branch stars by \cite{Renzini1987}.
Both \cite{Casanellas_Lopes13} and \cite{Casanellas+15} reported the effect in stars with $\sim 1.3 \Msun$ affected by ADM with varying $\sigxp$ and $\mx$ in a solar-like environment. The fixed values of ADM parameters used in this work are at the lower limits of the ranges considered there.
\cite{Lopes_2019} found similar suppression of convection in models of stars with $\Mstar \lesssim 2 \Msun$ in the Milky Way's nuclear star cluster ($\gb \sim 10^3$), with ADM properties $\sigxp=10^{-37} \cmsq$ and $\mx=4 \gev$.
Here we have explored the full range of ADM environments likely to be found in nature, and the full range of stellar masses that would be impacted. We show that not only is convection suppressed in the core, it moves into a shell that retreats away from the center as the star evolves. In the most extreme environments, stars with masses up to $\sim 4 \Msun$ are affected. 
See \S~\ref{sub:mstau} for further discussion of convective cores (and the impact on stellar lifetimes) with respect to varying stellar masses and environments.

\begin{figure*}
  \centering
  \includegraphics[width=\textwidth]{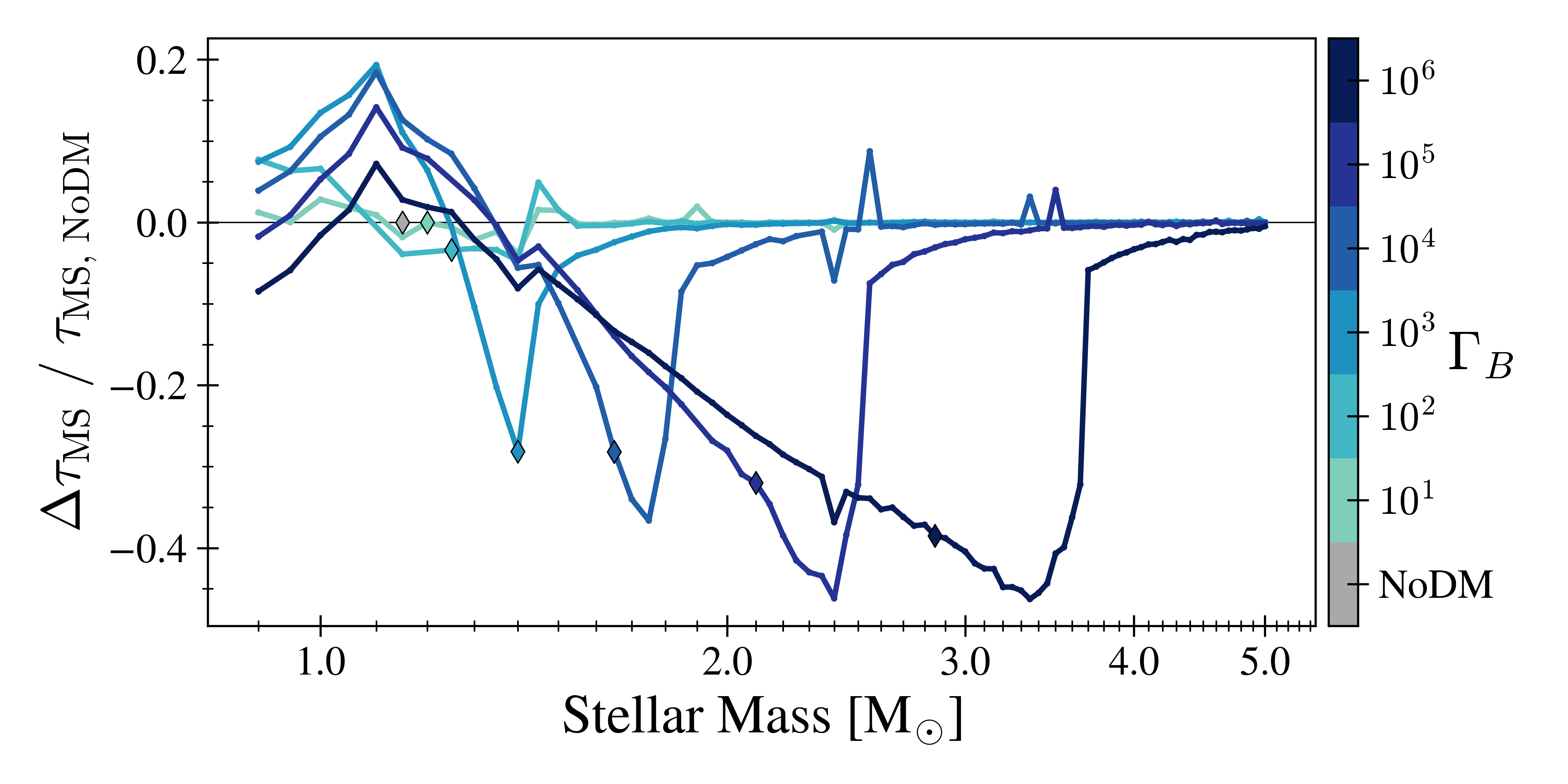}
  \caption{Changes in MS lifetimes, relative to a star of the same mass with no dark matter, seen in our simulations.
  Diamonds mark the transition from radiative to convective cores (left to right). For the purposes of this figure this is defined as the lowest $\Mstar$ for which the average (over the MS) mass of the convective core is greater than 0.01 Mstar. Stars to the right of the \nodm marker (grey diamond) have decreased lifetimes due to a reduction in the size of the convective core, which reduces the amount of hydrogen available for burning. The effect abruptly disappears as stellar lifetimes become shorter than the time required to build up a sufficient amount of ADM. Stars to the left of this marker show mixed behavior due to the competing effects of decreased central burning rates and higher temperatures around $m(<r) \gtrsim 0.1 \Msun$ which give the star access to more fuel. In addition to these trends, there are several abrupt dips (e.g., at $2.4 \Msun$) and spikes (e.g., at $2.55 \Msun$). This is due to rotational mixing that turns on part-way through the MS and funnels fresh hydrogen fuel to the center, which increases the lifetimes (spikes). The dips result when the \nodm model exhibits this feature, but the ADM model of the same mass does not.
  }
  \label{fig:mstau}
\end{figure*}

\subsection{Main Sequence Lifetimes}
\label{sub:mstau}

In Figure~\ref{fig:mstau} we summarize the effects of ADM on main sequence (MS) lifetimes relative to a standard \nodm star of the same mass. For the purposes of this paper, we have {\em defined} the MS to end when the fractional abundance of hydrogen in the center, $X_c$, falls below $10^{-3}$. Once $X_c < 10^{-3}$ the hydrogen burning rate is greatly reduced and the star transitions out of the MS and onto the sub-giant branch. This transition period is marked by relatively sudden and dramatic changes to the star's structure. Stars that capture large amounts of ADM can have significantly different core structures at the end of the MS than their standard model counterparts, and these differences affect the stars' transition out of the MS, including the duration, in ways that are qualitatively different than ADM's effect on the MS itself. Therefore, different choices in the definition of when a star leaves the MS can affect the results. Our relatively moderate choice of $10^{-3}$ highlights changes in the core of the star during the bulk of the MS, rather than changes in the transition period between the MS and the sub-giant branch. We discuss ADM's affects on this transition period in \S~\ref{sub:tracks}.

The MS lifetimes of relatively 
low-mass stars (near $\sim 1 \Msun$) can be altered by up to 20\%, however the sense and degree of the shift is not a monotonic function of the strength of the 
dark matter effect, parameterized by $\gb$. This complicated dependence on the amount of 
captured ADM is due to the competition between increased burning rates and increased 
availability of burnable hydrogen fuel as discussed in \S~\ref{sub:lowmass}.

At higher masses, the influence of ADM on stellar lifetimes is clearer. 
ADM shortens the lifetimes of high mass stars ($\Mstar \gtrsim 1.3 \Msun$). 
In \nodm models, the central convection zone extends beyond the burning region, 
giving the star a source of fresh nuclear fuel as hydrogen from outside 
of the core is mixed into the center. 
Since ADM shuts off convection in the center, 
the star no longer gets this influx of fresh hydrogen. 
Consequently, the star has less fuel available to burn, 
and so it leaves the MS faster than the \nodm models. 
Note that the appearance of a convective core (diamonds, Fig~\ref{fig:mstau}) shifts 
to higher masses with increasing $\gb$ 
due to larger amounts of ADM which can carry larger energy fluxes. 
The effect disappears abruptly as $\Mstar$ increases because stellar 
lifetimes scale as $\Mstar^{\sim 2.5}$ and 
quickly become too short for a sufficient amount of 
ADM to build up
(recall that the ADM capture rate scales roughly linearly with $\Mstar$), 
while the luminosity of the star increases rapidly with mass 
(roughly, $L \propto \Mstar^{3.5}$), 
meaning that more energy must be transported 
in order to alter the stellar structure.

These changes in MS lifetimes are consistent with those seen in \cite{Lopes_2019} for stars in an environment similar to the Milky Way Nuclear Star Cluster (roughly $\gbpow{3}$). Here we extend the study to a wider range of environments, including the most extreme environments likely to be found in nature, and therefore we see effects over a wider range of stellar masses. We will further extend the analysis to the period of transition out of the MS in \S~\ref{sub:tracks}.

In addition to the MS lifetime trends we have discussed, Figure~\ref{fig:mstau} has several abrupt dips (e.g., at $2.4 \Msun$) and spikes (e.g., at $2.55 \Msun$). This is due to rotational mixing that turns on part-way through the MS and funnels fresh hydrogen to the center, which increases the lifetimes (spikes). The dips result when the \nodm model exhibits this feature, but the ADM model of the same mass does not. This rotational mixing occurs sporadically (i.e. at isolated masses, not in a continuous range of masses) in our models and may or may not be physical. However, it cannot be a bug introduced by our ADM energy transport module since some \nodm models display the feature, but our module is not called in this case. This phenomenon where a star of a given mass receives an influx of hydrogen to the center due to the onset of mixing, while stars of bracketing masses experience no such mixing, has also been reported previously in the \mesa mailing lists\footnote{\url{https://lists.mesastar.org/pipermail/mesa-users/}}.

\subsection{Stellar Evolutionary Tracks}
\label{sub:tracks}

\begin{figure*}
  \centering
  \includegraphics[width=\textwidth]{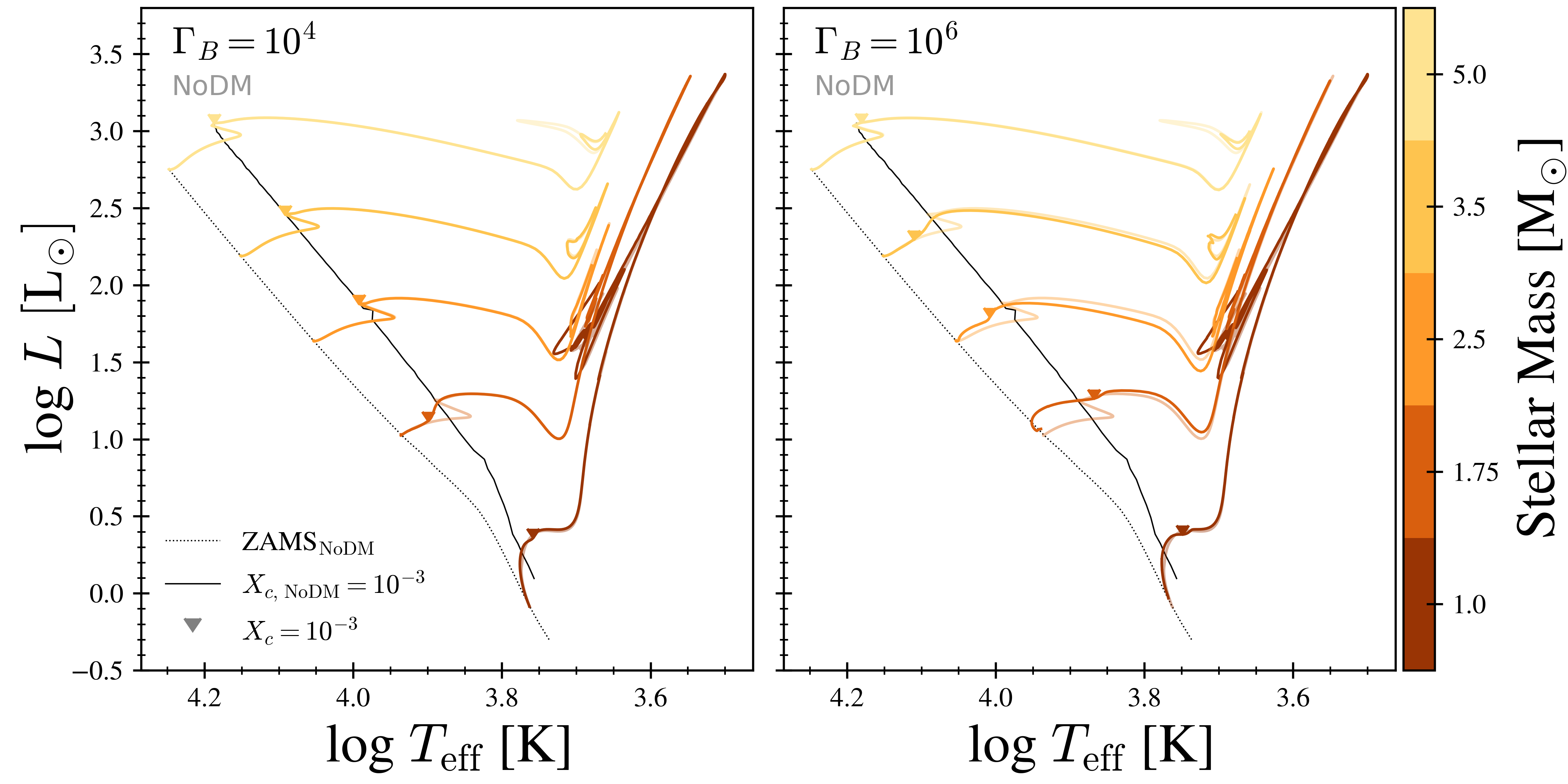}
  \caption{
  Stellar evolution tracks, from ZAMS to core helium depletion ($Y_c=10^{-3}$), of select masses with $\gbpow{4}$ (left) and $\gbpow{6}$ (right). Tracks of \nodm models of the same mass are overplotted with a higher transparency. The chosen masses highlight some of the most dramatic changes seen in Figure~\ref{fig:mstau}. Triangles mark the location where stars leave the MS, which we define here as core hydrogen depletion below $X_c=10^{-3}$. The location of the ZAMS and core hydrogen depletion for \nodm models are plotted as dotted and solid black lines respectively. The spike in the $X_{c,\ NoDM} < 10^{-3}$ line near the $2.5 \Msun$ track is due to rotational mixing in the $2.4 \Msun$ model, which is discussed in \S~\ref{sub:mstau}. The main effect of ADM on a star's surface properties is to move the star through roughly the same sequence of events at a faster or slower pace, causing the offset of the $X_c=10^{-3}$ milestone relative to \nodm. High mass stars with sufficient ADM skip the convective hook because ADM shuts off convection in the core. These stars transition into shell burning, and therefore the sub-giant branch, more smoothly, similar to low mass stars.
  }
  \label{fig:tracks}
\end{figure*}

\begin{figure*}
  \centering
  \includegraphics[width=\textwidth]{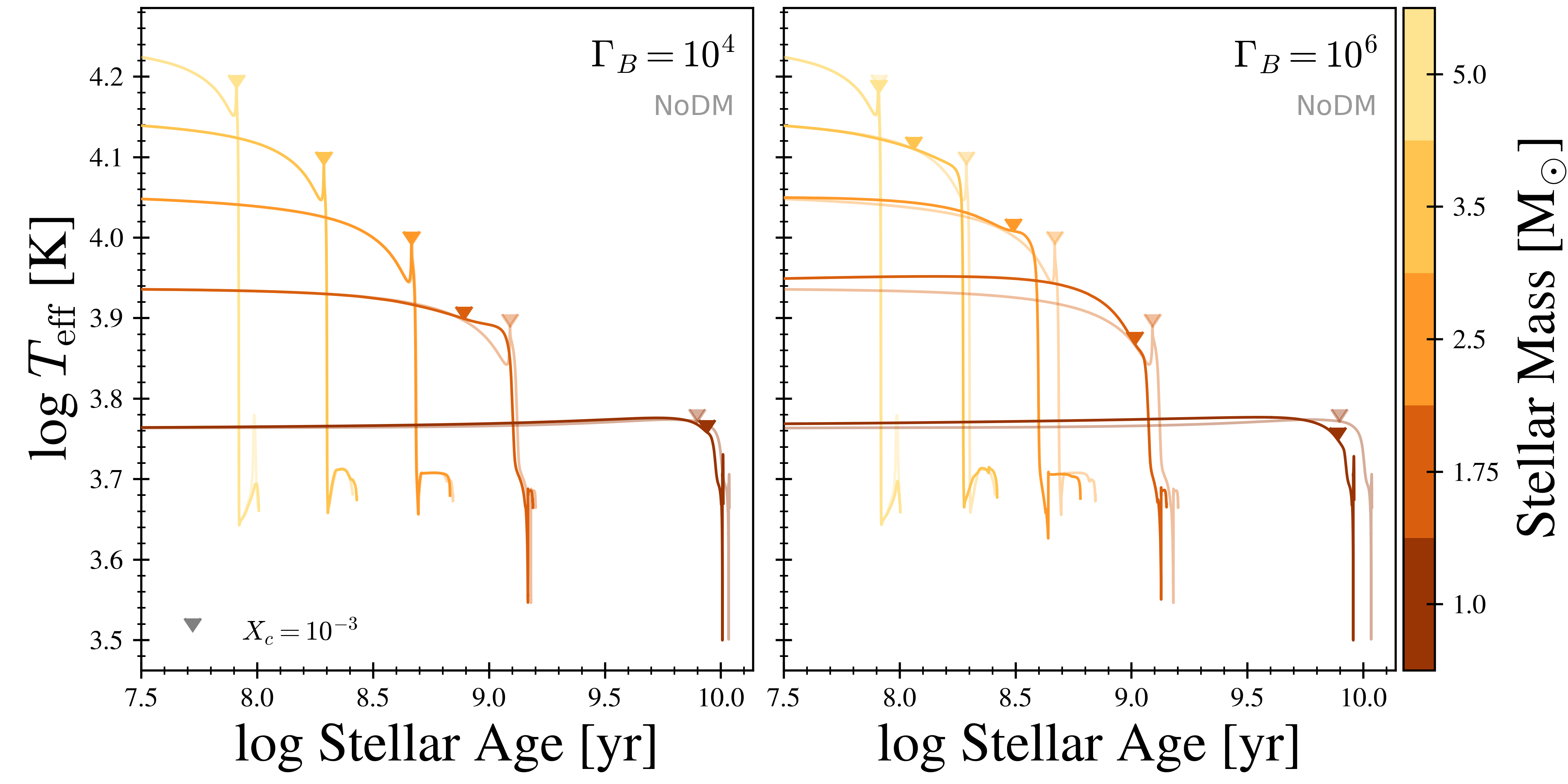}
  \caption{
  Same as Figure~\ref{fig:tracks}, except that here we plot effective temperature as a function of stellar age. The sub-giant branch phase in a star's evolution is seen here as a sharp drop in $\Teff$, and is a result of structural changes in the star that are triggered by large reductions in the core burning rate due to hydrogen depletion at the end of the MS. The duration of the transition period between the MS and the sub-giant branch is seen here as the temporal difference between the locations of $X_c = 10^{-3}$ (triangles) and the drop in $\Teff$. ADM alters the duration of the transition period, tending to increase it in high mass stars and decrease it in low mass stars. This is opposite of ADM's effect on MS lifetimes. The net effect is that the feature in $\Teff$ always occurs either concurrently or earlier in the ADM model than in its \nodm counterpart.
  }
  \label{fig:teff}
\end{figure*}


One of the goals of this work is to determine whether or 
not ADM can cause any gross changes to the properties of 
stars. We begin to answer this question with 
Figure~\ref{fig:tracks}, 
which shows evolutionary tracks on the 
HR diagram for many of our models. The tracks begin on the 
zero-age main sequence (ZAMS), delineated by the dotted 
black lines at the lower left of each panel. 
Stars evolve off of the ZAMS in a mass-dependent manner that is 
familiar from well-known aspects of standard stellar evolution. 
The tracks that we show leave the MS, defined as $X_{\rm c} < 10^{-3}$, 
at the points marked by triangles. The points at which stars exit  
the MS in a standard model with no dark matter are indicated by 
the solid black lines in each panel. (The spike near the $2.5 \Msun$ track is due to rotational mixing in the $2.4 \Msun$ model, which is discussed in \S~\ref{sub:mstau}.) Stars spend the majority of 
their lives on the MS and move more rapidly through the subsequent 
phases of stellar evolution. Our evolutionary tracks terminate 
when the core helium fraction falls below $10^{-3}$.

As is evident in Fig.~\ref{fig:tracks}, 
the effects of ADM on the evolutionary track 
of any individual star are generally subtle. 
Roughly speaking, this is not surprising.  
At a fixed central hydrogen fraction, stars containing ADM have different surface luminosities than their standard model counterparts, which is a result of ADM altering the structure of the star. This can be seen in the difference in location between the $X_c=10^{-3}$ markers in Fig.~\ref{fig:tracks} (x's for ADM models, solid black line for \nodm models).
However, if we instead consider stars at fixed luminosity, the temperature profiles, chemical compositions, 
opacities, and other properties of the overlying zones 
are approximately unaltered by ADM cooling. Consequently, 
the gross properties of the stellar photosphere, which are 
determined via the equations of stellar structure, are 
approximately fixed, at fixed luminosity. 
The result is that both ADM and standard model stars of a given mass follow roughly the same tracks in the HR diagram, but they do so at different rates.

Nonetheless, there are some small differences between 
standard evolutionary tracks and the tracks of stars 
with ADM. For example, consider the track of the 
$1.75 \Msun$ star in the left panel, corresponding 
to an environmental factor of $\gb = 10^4$.
The standard 
model of stellar evolution shows a kink in the evolutionary 
track as the star exits the MS. This kink is known as the 
{\em convective hook}. The convective hook is caused by 
an overall contraction of the convective cores of the 
stars after hydrogen depletion. During this phase, 
$T_{\rm eff}$ increases. Eventually, at the hottest point 
on the hook, contraction of the former convective core 
is sufficient to ignite burning in a shell. After this 
point, shell burning ensues and the star continues to 
evolve along the sub-giant branch. What is clear from 
the evolutionary track of the $1.75M_{\odot}$ star 
in the left panel of Fig.~\ref{fig:tracks} is that 
this evolutionary track exhibits {\em no convective hook.} 
This is because convection within the stellar core has 
been shut off by the ADM in these models. Instead of 
going through a phase of core collapse followed by shell 
burning, such stars make a smooth transition to shell burning 
and, thus, a smooth transition to the sub-giant branch. 
The absence of convective hooks is evident 
for a wider range of masses in the right hand 
panel of Fig.~\ref{fig:tracks}, which corresponds 
to a larger environmental factor of $\gb=10^6$. 


The convective hook feature has been clearly 
seen in many open clusters for which 
the main sequence turn off lies between $\sim 1.3\Msun$ 
and $\sim 2\Msun$, corresponding to stellar 
ages of $\sim 1\, \mathrm{Gyr}$ to $\sim 4\mathrm{Gyr}$ \citep[e.g., see Fig. 18 in][]{Gaia2018}. 
However, this does not yet provide any strong 
statement about the nature of dark matter because 
none of those environments are thought to contain 
significant amounts of dark matter. 
If a stellar population were identified within
the appropriate age range, associated with a significant amount of dark matter, 
and containing a sufficiently
large number of stars, the presence or absence of a convective hook should be immediately clear on an HR diagram.



In Figure~\ref{fig:teff} we plot the effective temperatures with respect to stellar age to better understand the transition from the MS to the sub-giant branch, seen here as the difference between $X_c = 10^{-3}$ (triangles) and the sharp drop in $\Teff$. A star's exit off of the main sequence is triggered when the core hydrogen fuel supply is depleted and the burning rate decreases such that it no longer provides sufficient pressure support and the star begins to collapse. Densities and temperatures increase until the bottom layer of hydrogen, now in a shell surrounding the core, ignites. The outward pressure resulting from increased shell burning causes the star's outer envelope to expand and cool at roughly constant luminosity, seen in Figure~\ref{fig:teff} as a large, sudden drop in $\Teff$. In models with no ADM, this transition period is more abrupt in high mass stars due to the mixing induced by their convective cores, which causes hydrogen to become depleted throughout the core simultaneously and shell burning to appear suddenly (see \S~\ref{sub:highmass}). In standard model low mass stars the cores are not convective, and so hydrogen depletes first at the very center and the burning shifts into a shell more gradually (see \S~\ref{sub:lowmass}). 

The temporal difference between ADM and standard models in the location of the sudden drop in $\Teff$ (when increased shell burning causes the envelope to expand) is another indicator of the change in MS lifetime. Unlike our definition of the end of the main sequence ($X_c = 10^{-3}$) this indicator is based on surface properties and occurs towards the end of the large structural changes the happen during the transition period. In some cases (e.g., in the $1.75 \Msun$, $\gbpow{4}$ model) this temporal difference is much smaller than the change in MS lifetime given by our definition of leaving the MS (seen here as the difference between the triangle markers of the ADM model and its standard model counterpart), and in other cases (e.g., in the $1.0 \Msun$, $\gbpow{6}$ model) it is larger.

ADM can affect both burning rates and stellar structure (e.g., convection), and therefore it is not surprising that ADM affects the timescale of a star's transition off of the main sequence. High mass stars that skip the convective hook due to ADM energy transport take longer to move through this transition period because the burning shifts gradually into a shell (see \S~\ref{sub:highmass} for details). This behavior is very similar to standard model low mass stars. Conversely, the $1.0 \Msun$ ADM models move through this period more quickly than their \nodm counterpart. This is likely due to the fact that ADM has caused higher burning rates at the outer edge of the core during the MS, so that mixing during this transition period brings more helium into the center than in the standard model star. These shifts are opposite the shifts in MS lifetimes, and the net result is that this feature in $\Teff$ always occurs either concurrently or earlier in the ADM model than in its standard model counterpart.

\subsection{Stellar Population Isochrones}
\label{sub:isochrones}

\begin{figure*}
\centering
   \begin{subfigure}{0.49\linewidth} \centering
     \includegraphics[scale=0.4]{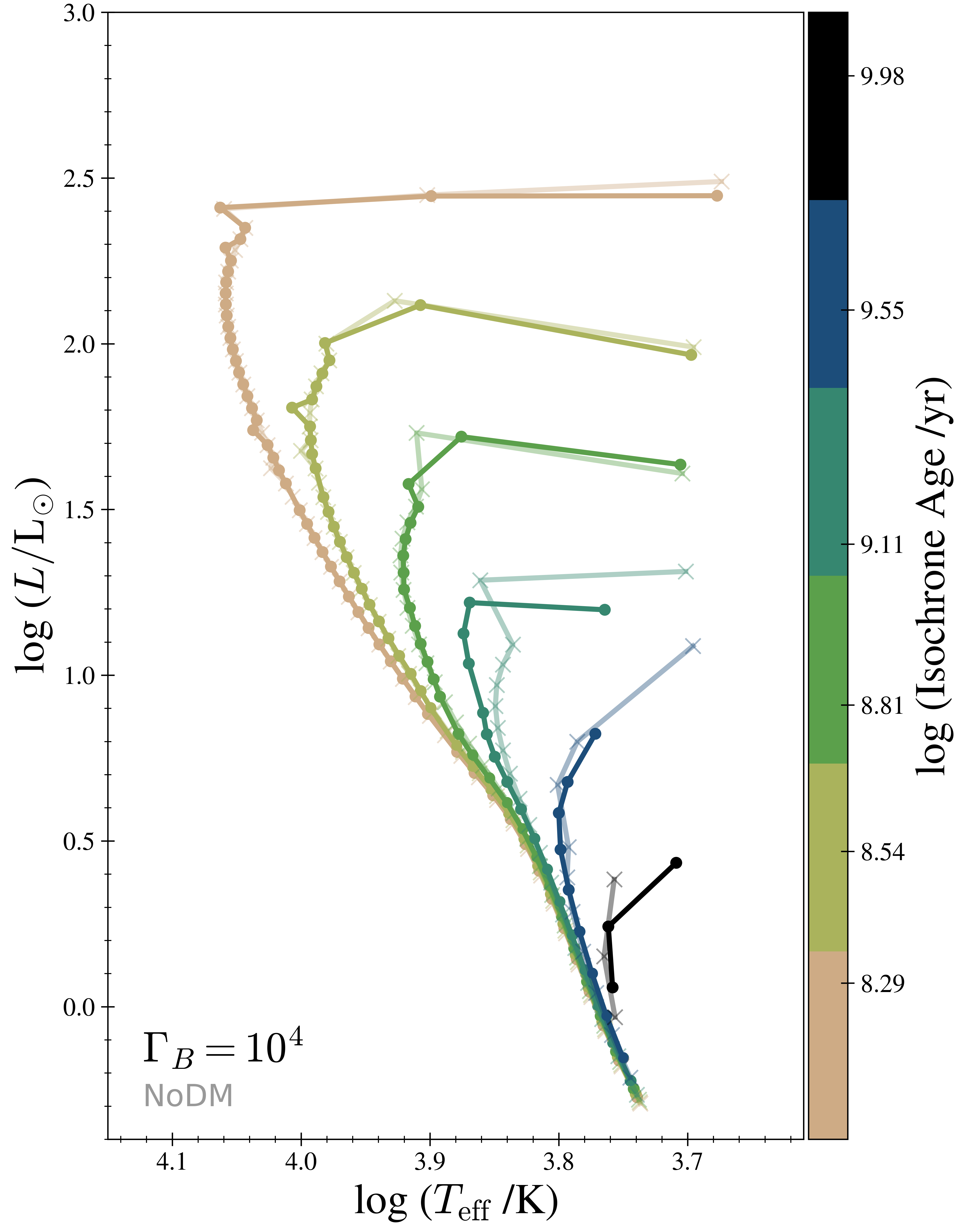}
   \end{subfigure}
   \begin{subfigure}{0.49\linewidth} \centering
     \includegraphics[scale=0.4]{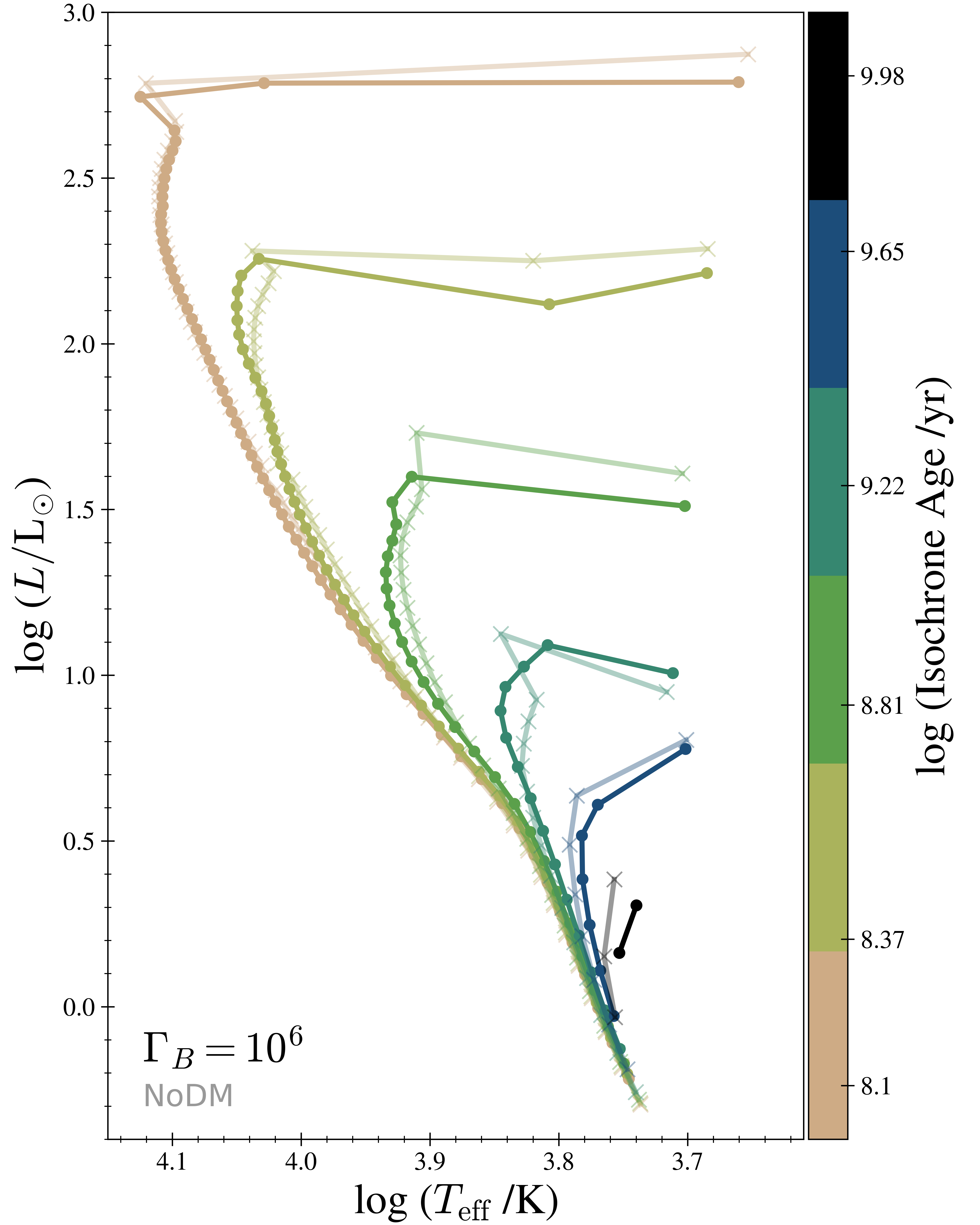}
   \end{subfigure}
\caption{Isochrones for $\gbpow{4}$ (left) and $\gbpow{6}$ (right) models, marked by circles, with \nodm models overplotted at higher transparency and marked by crosses. The data points are interpolated from the stellar tracks to a common set of times (isochrone ages). We connect the interpolated data points of a single isochrone with straight lines to guide the eye. The lowest data point on every isochrone is a $0.9 \Msun$ star. As stars leave the main sequence they evolve rapidly, and therefore subsequent phases are less well sampled due to the mass resolution. Isochrone ages have been chosen to maximize the sampling around the MS turnoff and sub-giant branch, and are not the same in each panel. We do not show the giant branches because we do not have enough data points there to be representative.
  The MS turnoff of isochrones around 1 Gyr happens at a higher effective temperature and a lower luminosity and skips the convective hook. This happens in a wider range of ages in populations that live in richer ADM environments (right panel). The oldest isochrones contain only low mass stars, and the ADM and \nodm populations look very similar except that populations in high $\gb$ environments appear slightly older due to their decreased luminosity and temperature.
  } 
\label{fig:isos}
\end{figure*}

\begin{figure*}
    \centering
    \includegraphics[width=\textwidth]{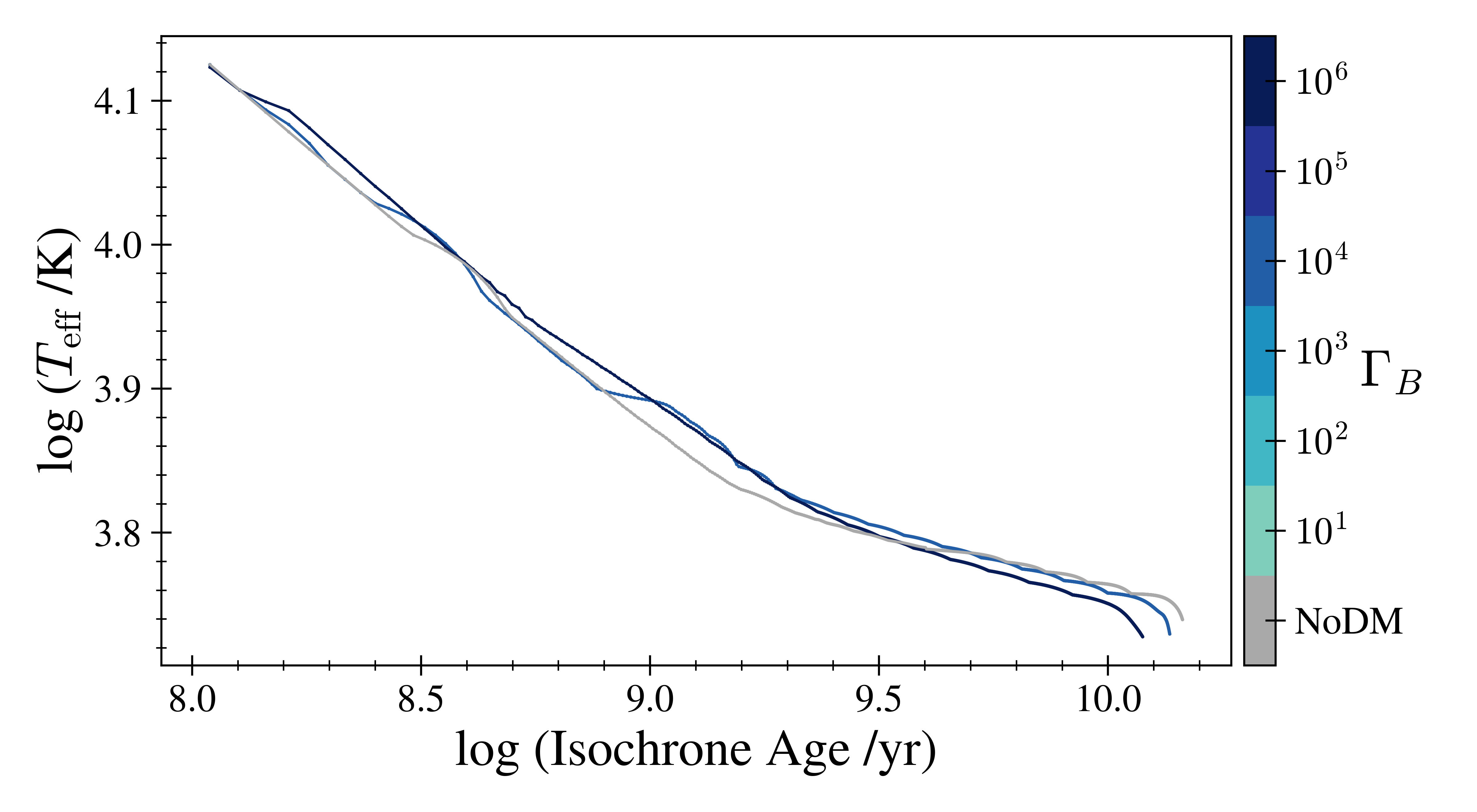}
    \caption{Effective temperature of the main-sequence-turnoff star, defined here as the hottest MS star, at a given age.
    We show only the \nodm, $\gbpow{4}$, and $\gbpow{6}$ models to allow the reader to see them more clearly. The MS turnoff temperature of the $\gbpow{4}$ models starts to become hotter than in the \nodm isochrones at $\sim 0.7$ Gyr when stars of $\sim 2 \Msun$ begin leaving the MS, while those of the $\gbpow{6}$ models become hotter at $\sim 0.1$ Gyr when stars of $\sim 3.5 \Msun$ begin leaving the MS. As we move to older isochrones the effect is reversed, and ADM models have lower effective temperatures at the turnoff. The lines terminate when there are no more stars on the main sequence (the lowest mass in our set of models is $0.9 \Msun$).
    For the purposes of this figure, we exclude stars in the convective hook because this feature is not well-resolved in our isochrones (see Fig.~\ref{fig:isos}).
    }
    \label{fig:hotTeff}

\end{figure*}

\begin{figure}
    \centering
    \includegraphics[width=0.45\textwidth]{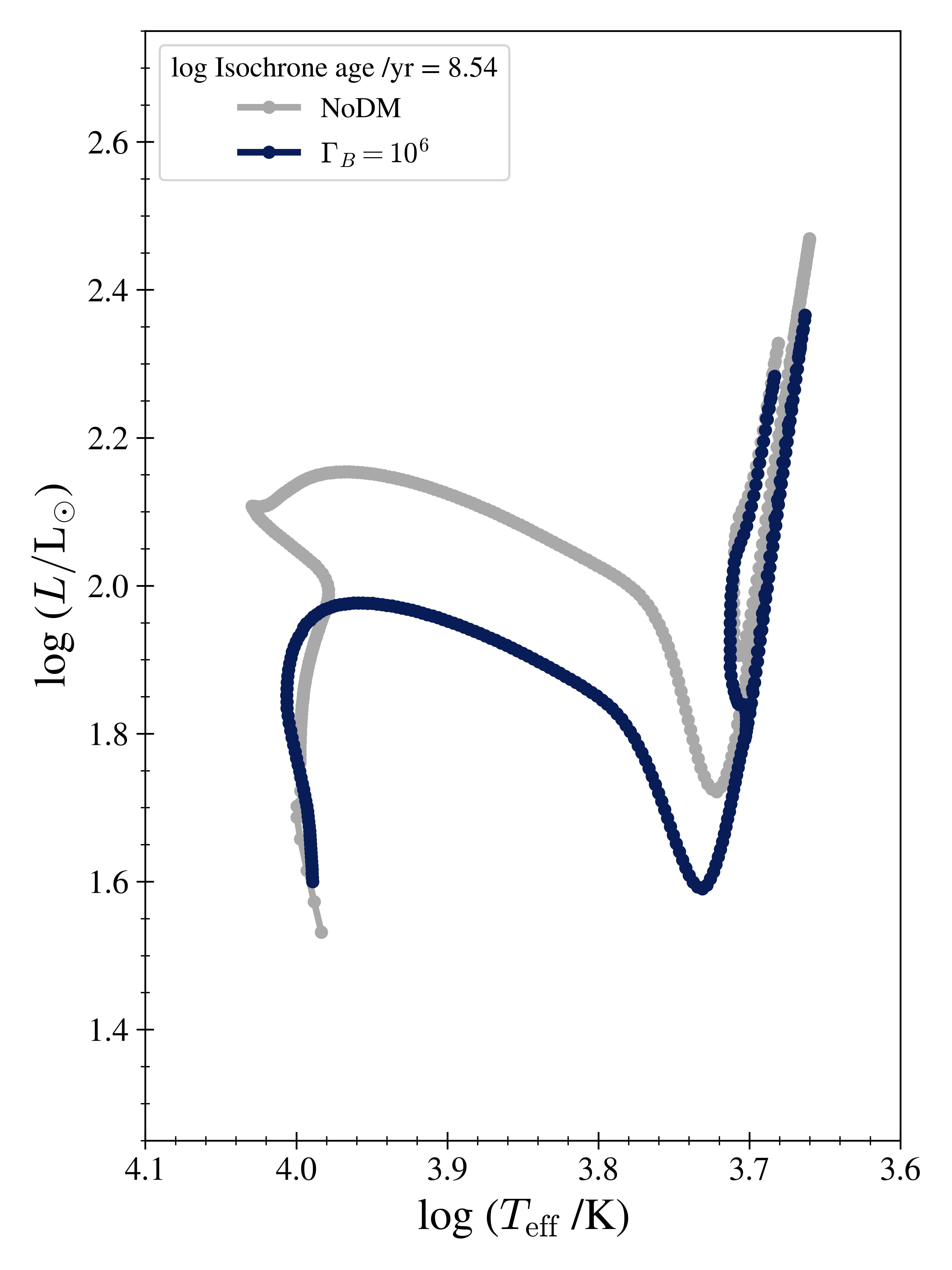}
    \caption{Isochrones generated by the \mist code for \nodm (grey) and $\gbpow{6}$ (dark blue) populations. The lowest mass star in both isochrones has $\Mstar = 2.25 \Msun$ (the interpolation was not successful for lower masses), and we show data through core helium depletion ($Y_c = 10^{-3}$) which corresponds to $\Mstar = 3.2 \Msun$ in both cases.
    The $\gbpow{6}$ isochrone skips the convective hook and crosses the sub-giant branch at a lower luminosity, consistent with Figure~\ref{fig:isos}. Additionally, the tip of its red giant branch occurs at a lower luminosity, a trend which appears in most of the $\gbpow{6}$ \mist isochrones.}
    \label{fig:isoDot}
\end{figure}

Though the evolutionary tracks are quite similar across all models, 
ADM changes the rate of evolution and, hence, the stellar ages at 
which stars reach particular evolutionary stages. To convey some 
of the information that is obscured in an evolutionary track, 
we present stellar isochrones in Figure~\ref{fig:isos}. Each isochrone is a line on this plot 
that represents the locus of points that would be occupied by a 
population of stars of fixed age, but a wide range of masses. 

The changes caused by ADM to individual stars' MS evolution is seen in these populations as a shift in the location of the MS turn-off, where the isochrones take a hard right turn. We have chosen these particular ages to maximize the sampling around this period and the subsequent crossing of the sub-giant branch. We do not show the giant branches because we do not have enough data points there to be representative, but we discuss the red giant branch further below. Due to the fact that stars move through the MS turnoff, sub-giant, and giant branches rather quickly, our mass sampling limits our ability to resolve these phases.

At around 1 Gyr, stars of $\Mstar \sim 1.75 \Msun$ are leaving the MS. In Figure~\ref{fig:isos}, we see that the MS turnoffs around this time occur at a higher effective temperature and skip the convective hook \citep[consistent with][]{Lopes_2019}, reflecting ADM's affects on these stars, discussed in \S~\ref{sub:highmass}. These isochrones also tend to cross the sub-giant branch at a lower luminosity. This reflects the fact that ADM speeds up the evolution of these stars, and so stars with smaller initial masses, which have lower effective temperatures, are crossing the sub-giant branch earlier than they otherwise would. This happens in a wider range of ages in populations that live in richer ADM environments (right panel). Stars move through this phase very quickly, meaning isochrones of real stellar populations are very sparsely populated in this region (known as the Hertzsprung Gap), however the location of the gap itself may contribute to discerning between models.

The oldest isochrones contain only low mass stars, since higher mass stars have already evolved into the giant branches and beyond. Here, the ADM isochrones look very similar to their standard model counterparts, except that populations in high $\gb$ environments appear slightly older due to their decreased luminosity and temperature. This indicates that ADM causes the stars' surface properties to evolve more quickly, likely due to the increase in shell burning.

To better resolve the isochrone's MS turnoffs, Figure~\ref{fig:hotTeff} shows the effective temperature of the MS turnoff star, which we define as the hottest MS star at a given age. At younger ages there is no difference between the ADM and standard models because the stars have not yet captured enough ADM to be significantly affected. Around 0.15 Gyr, isochrones of the $\gbpow{6}$ model start to display higher temperatures, remaining high until $\sim 3$ Gyr, after which their temperatures are cooler than their \nodm counterparts. The $\gbpow{4}$ isochrones show similar trends, but they occur at later times, since it takes longer for stars in lower $\gb$ environments to build up sufficient ADM. The waviness in the lines at older ages is a result of limited mass resolution.

The trends seen in our isochrones are consistent with what we were able to see in the isochrones generated by the \mist code, in regions where that code's interpolation was successful. (The reader is reminded that this method did not produce any isochrones older than 1 Gyr; see \S~\ref{sec:methods} for more information.) In addition, we noticed from \mist isochrones that the tip of the red giant branch tends to occur at a lower luminosity in populations with large amounts of ADM. The tip of the red giant branch is commonly used as a distance indicator, particularly for older populations. If the trend continues in isochrones older than 1 Gyr, ADM may add a source of uncertainty to these studies. 
To give the reader a sense of this shift, and to show a more well resolved MS turnoff and sub-giant branch, we show one particularly successful \mist isochrone for \nodm and $\gbpow{6}$ models in Figure~\ref{fig:isoDot}. The lowest mass star in both isochrones (lower left) has $\Mstar = 2.25 \Msun$ (the interpolation was not successful for lower masses), and we show data through core helium depletion ($Y_c = 10^{-3}$) which corresponds to $\Mstar = 3.2 \Msun$ in both cases.

\section{Discussion and Conclusions}
\label{sec:discus}


We have studied the potential impact of asymmetric dark matter interacting 
with nucleons through a spin-dependent coupling on the gross evolution 
of stars. We accomplished this by incorporating a module that approximates 
heat transport by dark matter into the {\tt MESA} stellar evolution 
software. We have identified several interesting qualitative distinctions between the standard evolution of stars and the evolution of stars in environments with a very high dark matter content. These include:
\begin{enumerate}[1)]
\item Flattened core temperature gradients, which alters the burning rates and stellar structures of low and high mass stars (where low mass stars have radiative cores and high mass stars have convective cores in models with no DM) in qualitatively different ways (\S~\ref{sub:lowmass} and \S~\ref{sub:highmass} respectively).
\item Convection is suppressed in the cores of high mass stars and pushed into a shell that retreats from the center (\S~\ref{sub:highmass}), resulting in the absence of a convective hook in stellar tracks (\S~\ref{sub:tracks}) and population isochrones (\S~\ref{sub:isochrones}).
\item Changes to MS lifetimes (defined here as $X_c > 10^{-3}$, \S~\ref{sub:mstau}).
Lifetimes of low mass stars are \emph{increased} by as much as 20\%. Lifetimes of high mass stars (\mrangehigh) are \emph{reduced} by as much as 40\%;
\item Stars in both mass regimes cross the sub-giant branch at younger ages (\S~\ref{sub:tracks}) and may reach the tip of the red giant branch at lower luminosities (\S~\ref{sub:isochrones}).
\end{enumerate}
Our results are consistent with previous work that considered similar ADM properties using a variety of stellar evolution codes, and we extend the field by considering the full range of ADM environments likely to exist in the universe and the range of affected stellar masses.

Finally, we find that strict energy conservation criteria in the stellar simulation code is crucial for a proper accounting of the effects on low mass stars with large amounts of ADM so as not to trigger large, nonphysical, self-reinforcing oscillations throughout the star (\S~\ref{sub:lowmass}).



It is interesting to speculate on ways in which these effects could be 
used to identify and/or constrain dark matter or ways in which these 
effects may, at least, serve as an element of uncertainty in the analysis 
of stellar populations. Any constraint on dark matter arising from these 
effects requires very high-quality observations of a stellar population
residing in an environment with a large ambient dark matter density and thus there will be a significant element of serendipity involved. 
Such a population could potentially be observed by the Rubin Observatory's Legacy Survey of Space and Time \citep[LSST,][]{2019ApJ...873..111I}, which is expected to observe hundreds of dwarf galaxies with very high mass/light ratios (spectroscopic followup would be required). The population would need to contain enough stars that the isochrone features we have identified are observable, and stars in the parameter space of interest that are bright enough for quality spectroscopic measurements.
If such a population is observed, our models suggest that if it is $\sim 1$ Gyr old, the hottest MS star should be slightly \emph{hotter} than expected for a population without ADM, and it should be slightly \emph{cooler} at $\sim 10$ Gyr (differences between \nodm and $\gbpow{6}$ models are of $\sim 5\%$ in both regimes).
Contemporary measurements of $\Teff$ regularly achieve precisions of a few percent, and can be as low as 1.5\% or lower with high resolution, high signal-to-noise \citep[see, for example,][and references therein]{2010A&A...515A.111S}.
In addition, the tip of the red giant branch may occur at a lower luminosity. The tip of the red giant branch is commonly used as a distance indicator
so ADM may add a source of uncertainty to these studies.
Finally, the metallicity is known to affect many of the properties we have discussed (e.g., the locations of various phases in the HR diagram), so ADM may be a contaminant here as well.

Future work along these lines includes 
\begin{inparaenum}[1)]
\item chemical abundance studies exploring the effects of altered core burning;
\item asteroseismology of Sun-like MS and red giant branch stars, which could be seen in (e.g.,) the small frequency separation--a diagnostic that is sensitive to the core structure of the star.
\item spin-independent ADM-nucleus scattering, which should have a larger effect during later phases when stars are burning helium.
\end{inparaenum}



\section*{Acknowledgements}

TJR thanks Brett Andrews for many helpful discussions, particularly in the writing process.
TJR has been supported by the Department of Physics and 
Astronomy and the Pittsburgh Particle Physics, 
Astrophysics, and Cosmology Center (PITT PACC) 
at the University of Pittsburgh, including as an Arts \& Sciences Fellow in Physics and Astronomy. 
TJH would like to acknowledge funding from the College of STEM at CSU-Pueblo.
ARZ was funded in part by the U.S. National Science 
Foundation (NSF) through grant NSF AST 1516266 and by the University of Pittsburgh. 
H.M.-R. was funded by the NASA ADAP grant NNX15AM03G S01. H.M.-R. also acknowledges support from a PITT PACC, a Zaccheus Daniel and a Kenneth P. Dietrich School of Arts \& Sciences Predoctoral Fellowship.

{\em Software}: 
\mesa \citep{paxton2010modules,Paxton2011ModulesMESA,Paxton2013,Paxton2015,Paxton2018,Paxton2019}, 
\mist \citep{Dotter2016MesaIsochrones, Choi2016MESAModels},
\texttt{MATPLOTLIB} \citep{Hunter2007},
\texttt{PANDAS} \citep{jeff_reback_2020_3715232},
\texttt{NUMPY} \citep{vanwderwalt2011}.

\section*{Data Availability}
The code required to reproduce our results, and the data for the models highlighted in this paper are available at \url{https://zenodo.org/record/4064115}.
Additional data will be shared on reasonable request to the corresponding author.
The full code used in the production of this paper is available at \url{https://github.com/troyraen/DM-in-Stars}.



\bibliographystyle{mnras}
\bibliography{references}



\appendix



\bsp	

\label{lastpage}

\end{document}